\documentclass[]{iopart}
\usepackage{amssymb}
\usepackage{graphics}
\usepackage{cite}

\newcommand{\di}{\mathrm{d}}
\newcommand{\mrm}[1]{\mathrm{#1}}
\newcommand{\mc}[1]{\mathcal{#1}}
\newcommand{\az}{\mbox{\boldmath$\alpha$}_{az_abz_b}}
\newcommand{\ab}{{az_abz_b}}
\begin{document}
\title{Subthreshold antiproton production in pC, dC and $\alpha$C
	reactions}
%\author{H M\"uller\dag, V I Komarov\ddag}
\author{H M\"uller$^1$, V I Komarov$^2$}
\address{$^1$ Institut f\"ur Kern- und Hadronenphysik,
        Forschungszentrum Rossendorf, Postfach 510119, D-01314 Dresden,
        Germany}
\address{$^2$ Joint Institute for Nuclear Research, LNP, 141980 Dubna,
	Russia}
\ead{H.Mueller@fz-rossendorf.de}

\begin{abstract} Data from KEK on subthreshold $\bar{\mrm{p}}$ as well
  as on $\pi^\pm$ and $\mrm{K}^\pm$~production in pC, dC and $\alpha$C
  reactions  at energies between 3.5 and  12.0~A GeV are described for
  the first  time  within a unified  approach.  We  use a  model which
  considers a nuclear reaction as an incoherent sum over collisions of
  varying  numbers of projectile   and  target nucleons.  It   samples
  complete  events and thus  allows for the simultaneous consideration
  of all  final particles including the decay  products of the nuclear
  residues.  The  enormous  enhancement  of the  $\bar{\mrm{p}}$ cross
  section, as well as the moderate increase  of meson production in dC
  and $\alpha$C compared to  pC~reactions, is well reproduced.  In our
  approach, the observed enhancement near  the production threshold is
  mainly due to the contributions from the interactions of few-nucleon
  groups.
\end{abstract}
%\submitto{\JPG}

%\maketitle
%\date{\today}
%
\section{Introduction\label{intro}}

Subthreshold  particle production in  a nuclear reaction is understood
as production below the energy threshold of  the considered process in
a  free nucleon-nucleon   (NN) collision.   It   is   thus  a  nuclear
phenomenon which may be  explained by rather different assumptions  on
the  properties of nuclear   matter  and on the  interaction dynamics.
Hereby, it is  an open question to  what  extent subthreshold particle
production is governed by properties  of the nuclear ground state wave
function  and  to what extent  by  the dynamical properties of nuclear
matter, not reflected in the ground state description.  The problem is
far  from  a  final  solution   at present   and evidently requires  a
systematic study of    high-momentum transfer  processes, among   them
subthreshold particle production.

First    measurements  of subthreshold   $\bar{\mrm{p}}$~production in
proton-nucleus collisions had been carried out
\cite{chamberlain55,chamberlain56,elioff62,dorfan65a} a  few   decades
ago.  Then investigations in nucleus-nucleus collisions
\cite{baldin88,caroll89,shor89,schroeter93,schroeter94}   and     more    
recent         studies       of         proton-nucleus       reactions
\cite{lepikhin87} followed. In \cite{chiba93} for the first time 
light-ion-induced $\bar{\mrm{p}}$~production has been investigated and
an enormous enhancement  of subthreshold $\bar{\mrm{p}}$~production in
deuteron-nucleus reactions compared to proton-nucleus interactions has
been found. This fact has been  confirmed by the  final results of the
KEK  group~\cite{sugaya98}.   They measured  d-   and $\alpha$-induced
reactions  in the energy region  between  2.5 and 5.0~GeV/nucleon  and
observed at  3.5~GeV/nucleon $\bar{\mrm{p}}$~cross sections nearly two
and three orders of magnitude  larger than measured in  proton-nucleus
reactions.

Most of the  descriptions  proposed  so  far  are based  on  transport
calculations~\cite{li94a,batko91,huang92,cassing92,teis93,teis94,cassing94a,batko94,hernandez95,sibirtsev98},
thermodynamical considerations \cite{koch89,ko88,ko89,dyachenko00}  or
multi-particle  interactions  \cite{danielewicz90} and  are devoted to
proton and/or    heavy-ion     induced reactions.      By    measuring
light-ion-induced     subthreshold $\bar{\mrm{p}}$~production the  KEK
group~\cite{sugaya98} intended to  provide  an  experimental test  for
models in this transition region  between proton and heavy-ion-induced
reactions.       To    the  best     of   our     knowledge  only  one
paper~\cite{cassing94a} has  been  published  so far, which   compares
proton-    and   deuteron-induced   $\bar{\mrm{p}}$~production      at
subthreshold energies.

It   should be  stressed that   all  mentioned papers  on subthreshold
$\bar{\mrm{p}}$~production    consider    the  $\bar{\mrm{p}}$~spectra
without any relation to  results  concerning other reaction  channels.
At  KEK~\cite{sugaya98}, however,    the spectra   of  $\pi^\pm$   and
$\mrm{K}^\pm$~mesons  were  measured    together with    those of  the
antiprotons.  In a recent  paper~\cite{Komarov04}, in the framework of
the Rossendorf collision  (ROC) model we simultaneously considered all
these    reaction channels for  pC~reactions     and achieved a   good
reproduction  of   the   data.  In   this   paper,   we extend   these
considerations to dC and $\alpha$C reactions  using the same parameter
set.

In   the ROC model, it  is  assumed that  the  nuclear residue becomes
excited during  the reaction  due  to  the  distortion of the  nuclear
structure  by the separation of the   participants from the spectators
and due to the passage of the reaction  products through the spectator
system.  In this way, final-state interactions  are taken into account
without    making   special  assumptions  concerning    re-absorption,
re-scattering,  self-energies,  potentials etc for each  particle type
(see also \cite{Komarov04}).  The fragmentation of the residual nuclei
and the interaction  of  the nucleons participating in  the scattering
process are treated on the  basis of analogous assumptions.  Even more
important,  the phase-space of the  complete final state consisting of
the  reaction products of the interaction  of the participants as well
as  the  fragments    of  the  decay    of  the spectator  systems  is
calculated. exactly This feature seems to be unique to the ROC model.

The plan of the paper is as follows.  In section~\ref{model}, the main
ingredients of  the ROC model  are  explained, which is  used for  the
calculations  to    be  presented.    Section~\ref{comp}  contains   a
comparison  of   theoretical  and experimental   results  for particle
production  with   special  emphasis  on subthreshold  $\bar{\mrm{p}}$
production. A summary is given in section~\ref{sum}.

\section{The model\label{model}}
The ROC model is implemented as a Monte  Carlo generator which samples
complete  events  for hadronic as well   as  nuclear reactions. It was
successfully   tested for   pp  interactions up   to  ISR  energies in
\cite{muellerh95,muellerh01}, while  nuclear reactions were considered
in                          the                                 papers
\cite{Komarov04,muellerh92,muellerh91,muellerh95a,muellerh96}. A 
detailed  description of the model is  given  in \cite{muellerh01} for
the     case  of  pp interactions       and in  a   recent publication
\cite{Komarov04} for hadron-nucleus interactions. In the following the
basic mathematics   from the  hadron-nucleus  case \cite{Komarov04} is
extended  to  nucleus-nucleus reactions. Although    in this paper  we
consider only  light-ion-induced reactions, the given  presentation is
applicable to any nucleus-nucleus reaction.  For the interested reader
some minor changes in the ROC  model relative to previous publications
are indicated where appropriate.

The cross section of the interaction of two nuclei $(\mc{A},\mc{Z}_A)$
and $(\mc{B},\mc{Z}_B)$ characterized by their mass and charge numbers
is considered as an   incoherent sum over contributions  from  varying
numbers $a$ and $b$ of   nucleons  (thereof $z_a$ and $z_b$   protons)
participating in the interaction
\begin{eqnarray} \label{dsigma}
  \di\sigma(s)                      &                  =             &
  \sum_{a=1}^\mc{A}\sum_{z_a=max(0,a-\mc{N}_A)}^{min(a,\mc{Z}_A)} \,
  \sum_{b=1}^\mc{B}\sum_{z_b=max(0,b-\mc{N}_B)}^{min(b,\mc{Z}_B)}
  \nonumber \\ 
  &&\sigma_\ab \,\frac{\di W(s;\az)}{\sum_{\az}\int \di W(s;\az)}\,\,.
\end{eqnarray}
Here, $s=P^2$ denotes  the square of the  centre-of-mass energy of the
projectile-target     system with  $P=(E,\vec{P})$    being  the total
4-momentum.  The summation  extends over  all possible  charge  states
ranging  from  clusters  consisting completely   of neutrons  to those
composed totally of protons.  This  range is restricted in cases where
the  number of available protons  $\mc{Z}_A$  ($\mc{Z}_B$) or neutrons
$\mc{N}_A=\mc{A}-\mc{Z}_A$  ($\mc{N}_B=\mc{B}-\mc{Z}_B$)    is smaller
than the number of  cluster nucleons.  Expression $\sigma_\ab$  stands
for the partial cross section of the interaction of clusters $(a,z_a)$
with  $(b,z_b)$   , and  the  quantities  $\di W(s;\az)$  describe the
relative probabilities of the various final channels $\az$.

The partial cross    sections  $\sigma_\ab$ are  calculated  using   a
modified version of a Monte Carlo code~\cite{shmakov88} which is based
on    a    probabilistic     interpretation       of   the     Glauber
theory~\cite{glauber70}.  They  account    for   sequential collisions
between $a$  projectile  and $b$ target   nucleons  capable of sharing
their  energy  in   close analogy  with the   virtual  clusters of the
cooperative model
\cite{knoll79,knoll80,bohrmann81,shyam84,shyam86,knoll88,ghosh90,ghosh92}
(see \cite{Komarov04}  for a discussion of the  novel features  of the
ROC model  compared to  the   cooperative model). We  use the  profile
function
\begin{eqnarray}
\Gamma_{\mc{AB}}(d)&=&\int \left[1- \prod_{i=1}^\mc{A} 
   \prod_{j=1}^\mc{B} (1-p_{ij})\right] \nonumber \\
   &&  \prod_{i=1}^\mc{A} \rho_\mc{A}(\vec{r}_i) \di^3r_i
       \prod_{j=1}^\mc{B} \rho_\mc{B}(\vec{r}_j) \di^3r_j \nonumber
\end{eqnarray}
of  the considered nuclei,   which  depends on the nucleon   densities
$\rho_\mc{A}(\vec{r}_i)$, $\rho_\mc{B}(\vec{r}_j)$ and the probability
\[
   p_{ij}=\exp(-d_{ij}^2 \pi/\sigma_{NN})
\]
for an interaction  of  the  $i$th  projectile and the  $j$th   target
nucleon with $d_{ij}$  being   the distance between  the   interacting
particles.  The nucleon density \cite{elton61}
\begin{equation} \label{rho}
  \rho_\mc{A}(\vec{r})                                         \propto
  (1+\eta[1.5(f^2-e^2)/f^2+e^2r^2/f^4])\exp(-r^2/f^2)
\end{equation}
of light nuclei $2 < \mc{A} < 20$ can be derived from a standard shell
model wavefunction  with  $\eta=(\mc{A}-4)/6$ , $e^2=f^2/(1-1/\mc{A})$
and    $f=1.55\,\mrm{fm}$.  For   the   deuteron  the   Paris deuteron
wavefunction  \cite{lacombe81}  is used.   Then the   NN~cross section
$\sigma_{NN}$    is adapted such   that  the integral   of the profile
function over the impact parameter  $d$ reproduces the total inelastic
$\mc{AB}$ cross section
\[
\sigma_\mc{AB}^{in} = \int \di^2d\, \Gamma_\mc{AB}(d).
\]
The same     calculation  yields  also   the   partial  cross sections
$\sigma_\ab$   we are  interested    in   (for further  details   see
\cite{shmakov88}).

In (\ref{dsigma}), the quantities
\begin{equation} \label{dWaz}
  \di W(s;\az)   \propto  \di L_{n}   (s;\az)
  \rho_\mc{A}(\vec{P}_A) \rho_\mc{B}(\vec{P}_B) T^2
\end{equation}
describe the  relative  probabilities of   the various final  channels
$\az$.   They  are given by  the  Lorentz-invariant phase-space factor
$\di L_n(s;\az)$ multiplied by   the square of the  empirical reaction
matrix element $T^2$   describing the collision  dynamics.   The Fermi
motion is implemented via  the momentum distributions of the  residual
nuclei   $\rho_\mc{A}(\vec{P}_A)$ and $\rho_\mc{B}(\vec{P}_B)$,  which
are functions of  the numbers $a$  and $b$ of  participants.  They are
taken as Gaussians having, in case of nucleus $\mc{A}$, a width of
\begin{equation} \label{sigA}
   \sigma_a=\sqrt{aA/5/(\mc{A}-1)}\,\,p_F
\end{equation}
in accordance with  the  independent particle model~\cite{goldhaber74}
with  $A=\mc{A}-a$ being  the number of  nucleons in   the residue and
$p_F$ is  the Fermi limit   of the  nucleus considered.    For nucleus
$\mc{B}$ an analogous  formula holds, and  for the deuteron  again the
Paris  deuteron wavefunction  \cite{lacombe81}   is employed, now  in
momentum      space.             No           special    high-momentum
component~\cite{shor90,geaga80,ciofi96,benhar89,sick94,sibirtsev97} is
used in this paper.

The Lorentz-invariant phase-space is defined as  the integral over the
3-momenta  of the $n$    primarily produced final   particles with
energy-momentum conservation taken into account
\begin{equation} \label{phase-space1}
  \di  L_n(s;\az)   =  \prod_{i=1}^n\frac{\di^3p_i}{2e_i} \,  \delta^4
  (P-\sum_{i=1}^n p_i),
\end{equation}
Here,   the  4-momentum of   the    $i$th particle  is denoted  by
$p_i=(e_i,\vec{p}_i)$ with $p^2_i=m^2_i$.

For     numerical     calculations,      the     $\delta$-function  in
equation~(\ref{phase-space1}) has to be  removed by introducing a  new
set of $3n-4$ variables to replace the $3n$ 3-momentum components.  It
is reasonable to  choose  a  set of   variables, which  reflects   the
underlying  physical   picture   of the interaction   process.   Using
recursion
\cite{byckling73} equation~(\ref{dWaz}) can be rewritten in the form
\begin{eqnarray} \label{dWaz1}
   \di  W(s;\az)&   \propto &
   \frac{\di^3P_A}{2E_A} \rho_\mc{A}(\vec{P}_A) 
   \frac{\di^3P_B}{2E_B} \rho_\mc{B}(\vec{P}_B) \nonumber \\
   &&\di W_A(M_A^2) \di W_B(M_B^2) \di W_C(M_C^2)
\end{eqnarray}
with the   4-momenta  of the nuclear  residues  $P_A=(E_A,\vec{P}_A)$,
$P_B=(E_B,\vec{P}_B)$ and of the participants
\begin{equation} \label{dPP}
P_C=P-P_A-P_B=(E_C,\vec{P}_C),
\end{equation}
the  corresponding   invariant  masses are   given  by  $M_A^2=P_A^2$,
$M_B^2=P_B^2$ and $M_C^2=P_C^2$.  The integrals  over the Fermi motion
$\di^3P_A         \rho_\mc{A}(\vec{P}_A)/2E_A$        and    $\di^3P_B
\rho_\mc{B}(\vec{P}_B)/2E_B$     separate,  in  accordance  with    the
participant-spectator picture, the phase-spaces of the $n_A$ and $n_B$
nuclear fragments
\begin{equation} \label{dWR}
 \fl  \di W_A(M_A^2)= \di   M_A^2  \di L_{n_A}(M_A^2) T_A^2
   \quad \quad
   \di W_B(M_B^2)= \di   M_B^2  \di L_{n_B}(M_B^2) T_B^2
\end{equation}
from  the   phase-space 
\begin{equation} \label{dWC}
   \di W_C(M_C^2) = \di L_{n_C}(M_C^2) T_C^2
\end{equation}
of the  $n_C=n-n_A-n_B$  final particles arising from  the participant
system. In (\ref{dWR}) and (\ref{dWC}), the  matrix element $T^2= T_A^2
T_B^2 T_C^2$ is split  into  factors describing residue  fragmentation
($T_A^2$ and  $T_B^2$) and  participant interaction ($T_C^2$).   There
is, however,  a   strong   kinematic link   between  participants  and
spectators, since  invariant  mass $M_C$  of the  participant  system,
invariant masses  $M_A$  and $M_B$ of   the residues and the  relative
kinetic energy $\sqrt{s}-M_A-M_B-M_C$ of  these three particle  groups
are connected by  energy-momentum conservation.  The excitation energy
of  the  target  residue comes usually   into   play via the  spectral
function (see  e.g. \cite{ciofi96})  derived from  electron scattering
data.  Unique to   the ROC model,   however, is the   treatment of the
spectator system  in  close  analogy with the   participant subsystem,
since it calculates the complete final  state of both the participants
and the  spectators.

The term $\di  W_C(M_C^2)$ in (\ref{dWaz1}) and (\ref{dWC})  describes
the interaction  of  the groups of    participating nucleons.  Such  a
cluster-cluster reaction is treated in  complete analogy to a hadronic
reaction.   In    a first  step   intermediate  particle groups called
fireballs  (FBs)  are  produced, which decay    into so-called primary
particles.  The primary particles define  the  channels for which  the
weights (\ref{dWC}) are  calculated.  Among them are resonances, which
decay  subsequently into stable hadrons.   The  dynamical input of the
cluster-cluster reaction  is  implemented by  the empirical transition
matrix element

\begin{equation} \label{A2P}
  T^2_C= T^2_{\mrm i} T^2_{\mrm {qs}} T^2_{\mrm {ex}} T^2_{\mrm
  t} T^2_{\mrm l} T^2_{\mrm {st}}\,,
\end{equation}
which describes the  interaction  process $T^2_{\mrm i}$ resulting  in
the production of $N \ge  2$~FBs, the production of hadrons $T^2_{\mrm
{qs}}$ via  the creation of  quark-anti-quark ($q\bar{q}$) pairs , the
invariant-mass distribution of the FBs $T^2_{\mrm ex}$, the transverse
$T^2_{\mrm  t}$ and longitudinal  $T^2_{\mrm l}$ momentum distribution
of the FBs, and finally, some factors  $T^2_{\mrm {st}}$ necessary for
the  calculation  of  the  statistical weights.    The  interaction is
assumed to  proceed via  colour exchange   leading  to the  removal of
valence quarks or of gluons from the  interacting hadrons.  Additional
up, down and strange quark pairs are created in the ratio
\begin{equation} \label{lamba}
u:d:s = 1:1:\lambda_s
\end{equation}
with  $\lambda_s=0.15$.  They form  the varying  number of  FBs, which
subsequently  decay into  the final  hadrons.  The  transverse momenta
$P_{\mrm  t}$ of  the FBs  are  restricted by  an exponential  damping
(longitudinal phase-space) according to
\begin{equation} \label{At}
  T^2_{\mrm t} = \prod_{I=1}^N \exp(-\gamma P_{\mrm {t,I}})
\end{equation}
with  the  mean $\bar{\mrm{p}}_{\mrm  t}=2/\gamma$.   Two  leading FBs,  the
remnants of the incoming clusters, get in the mean larger longitudinal
momenta than the central FBs by weighting the events with
\begin{equation} \label{Al}
  T^2_{\mrm l} = (X_1 X_2)^\beta.
\end{equation}
Here, the   light-cone variables  $X_1=(E_1+P_{\mrm z,1})/(e_a+p_{\mrm
z,a})$ and $X_2=(E_2-P_{\mrm z,2})/(e_b-p_{\mrm  z,b})$ are  used with
the 4-momenta of the  participants given by  $p_a=(e_a,\vec{p}_a)$ and
$p_b=(e_b,\vec{p}_b)$.     In    \cite{muellerh01},  the   high-energy
approximations
\[
  X_1=(E_1+P_{\mrm  z,1})/\sqrt{s}   \quad       \mbox{and}      \quad
  X_2=(E_2-P_{\mrm z,2})/\sqrt{s}
\]
were  applied  for the  more  symmetric  pp interactions  without  any
substantial influence on  the results. For  completeness, it should be
mentioned that in the older publication \cite{muellerh95} a transition
matrix element with transverse damping only for the central FBs
\[
T^2_{\mrm t} = \prod_{I=3}^N \exp(-\gamma (P_{\mrm {t,I}})^2)
\]
was applied while  the 4-momenta $t_{a1}$ and $t_{b2}$  transferred
to the leading FBs were restricted via
\[
T^2_{\mrm l} = \exp{\left(\beta\left(t_{a1}+t_{b2}\right)\right)}.
\]
This variant worked well up to energies of about 50~GeV.

Each FB   is     characterized  by two   parameters,    a  temperature
$\Theta_{FB}$ and a  volume $V_{FB}$.  The temperature  determines the
relative kinetic energy of the particles the FB decays into via
\begin{equation} \label{AexP}
  T^2_{\mrm ex}(\Theta_{FB}) =
	\prod_{I=1}^N (M_I/\Theta_{FB})  K_1(M_I/\Theta_{FB}),
\end{equation}
while the volume defines  the interaction region and influences mainly
the particle multiplicity via  the statistical factor $T^2_{\mrm {st}}
\propto V_{FB}^{n_C-1}$. In (\ref{AexP}) $K_1$ stands for the modified
Bessel function. Final hadrons are built-up by random recombination of
the  available quarks  during the decay  of  the FBs.   This procedure
automatically  ensures    the  conservation of  all   internal quantum
numbers.  Subsequently,  resonances   decay   until the  final   state
consisting of stable particles  is    reached.  For a more    detailed
discussion of the hadronic matrix element the reader is referred to
\cite{muellerh01}.   In  this  paper, we use   the same  set of
parameters    as  in \cite{muellerh01} for  the     description of the
interaction of a single    projectile nucleon  with a single    target
nucleon.  Most of the other  terms in (\ref{dsigma}), however, contain
clusters consisting of several nucleons.   The basic parameters of the
FBs emerging from such an interaction are fixed  by scaling the volume
and the temperature parameter according to
\begin{equation} \label{sca}
V_{FB} = V_{FB}^0 \,(a+b-1) \quad \mbox{and} \quad
\Theta_{FB} = \Theta_{FB}^{max}\,(a+b-1)^{-1/3}.
\end{equation}

It remains to consider the target residues,  the structure of which is
strongly  disturbed  by   the  interaction  of  the   participants and
subsequent final-state interactions.  This leads to the excitation and
decay of  the spectator systems,  which  are characterized  by the two
parameters temperature   $\Theta_R$  and   volume $V_R$  ($R=A   \quad
\mbox{or} \quad B$ in dependence  on the residue under  consideration)
in the same way as the FBs emerging  from the participant interaction.
The part  of   the   matrix  element  responsible for    the   residue
fragmentation
\[
   T_R^2(\az^R)  = T^2_{\mrm   {ex}}(\Theta_R) T^2_{\mrm  {st}}(\az^R)
   \quad R=A,B
\]
is identical with the corresponding factors (\ref{A2P}) applied to the
hadronic FBs.  In order  to restrict the excitation energy transferred
to the residue we use the asymptotic approximation of (\ref{AexP}) for
large mass and small temperature
\begin{equation} \label{Aex}
  T^2_{\mrm   ex}(\Theta_R) = \sqrt{M_R/\Theta_R}  \exp(-M_R/\Theta_R)
  \quad R=A,B.
\end{equation}

An impact  parameter  dependence is    assumed  for the    temperature
parameter.   This seems    to be  reasonable,  because   a  peripheral
collision with only  few   participating nucleons should excite    the
nuclear  residue  much less  than   a   central collision  with   many
participants.  As a first guess we use
\begin{equation} \label{ThetaR}
\fl \Theta_A = \Theta_A^{max} [1-\exp{(-a/\bar{a}A^{1/3})} \quad \quad
\Theta_B = \Theta_B^{max} [1-\exp{(-b/\bar{a}B^{1/3})}
\end{equation}
with $\bar{a}$   as parameter,   here fixed to    $\bar{a}=0.5$, which
determines  how  fast the  maximal  temperatures  $\Theta_A^{max}$ and
$\Theta_B^{max}$ are reached  with increasing number of  participants.
The value $\Theta_A^{max}=\Theta_B^{max}=12\,\mrm{MeV}$ has been fixed
in \cite{Komarov04}   from  a comparison  of   the energy  spectra  of
fragments from the reaction of 2.1~GeV protons with carbon measured at
$\mrm{90}^\circ$ \cite{westfall78} with ROC-model calculations.

All factors still necessary for a  correct calculation of the relative
weights of the various channels are collected in complete analogy with
the corresponding factors for the FBs in the product of two terms with
$R=A$ and $R=B$ according to
\begin{equation} \label{Ast}
  T^2_{\mrm{st}}(\az^R)=g(\az^R)
  \left(\frac{V_R}{\left(2\pi\right)^3}\right)^{n_R-1}
  \prod_{i=1}^{n_R}(2\sigma_i+1)2m_i.
\end{equation}
It contains the spin degeneracy factors $(2\sigma_i+1)$ and the volume
$V_R$ in which the particles are produced.  The quantity $g(\az^R)$ is
the  degeneracy factor for groups  of identical particles in the final
state of the residue decay and prevents multiple counting of identical
states.  Further details  of    residue treatment  can  be found    in
\cite{Komarov04}.

\section{Comparison with experimental data \label{comp}}
\begin{figure}
\begin{center}
   \resizebox{0.98\textwidth}{!}{
   \includegraphics*{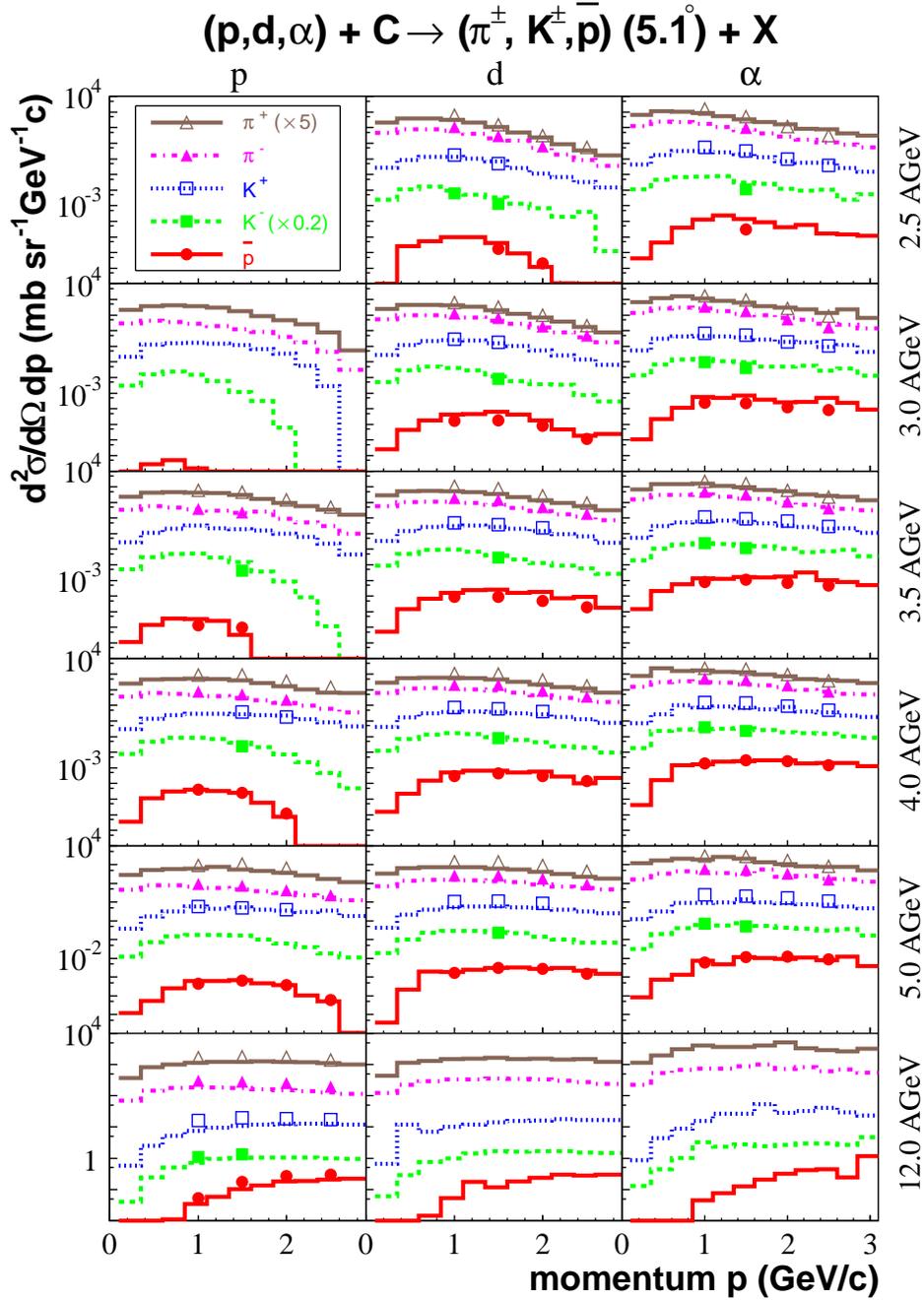}}
\end{center}
   \caption{\label{fCAll} Momentum spectra of $\pi^\pm$, $\mrm{K}^\pm$
   and  $\bar{\mrm{p}}$ \cite{sugaya98}   (symbols)  from pC, dC   and
   $\alpha$C reactions   compared with        ROC-model   calculations
   (histograms). Experimental and  calculated results for $\pi^+$  and
   $\mrm{K}^-$ mesons  are multiplied by the  factors indicated in the
   legend.   The scale extends from $10^{-8}$  to  $10^4 \, \mrm{mb \,
   sr^{-1} \,  GeV^{-1}\,c}$ for energies up  to 4.0~A GeV,  while the
   lower limit is  $10^{-6}$ for 5.0~A  GeV and $10^{-2} \, \mrm{mb \,
   sr^{-1}\, GeV^{-1}\,c}$ for 12.0~A GeV.}
\end{figure}

In  figure~\ref{fCAll},  the     momentum    spectra of     $\pi^\pm$,
$\mrm{K}^\pm$ and $\bar{\mrm{p}}$  from pC, dC and $\alpha$C reactions
measured   at  KEK~\cite{sugaya98}   are  displayed.  The experimental
efforts of the KEK group were motivated by measurements of large cross
sections of subthreshold $\bar{\mrm{p}}$~production in nucleus-nucleus
reactions   at    LBL-BEVALAC~\cite{caroll89,shor89}       and      at
GSI~\cite{schroeter93}.   The authors~\cite{sugaya98} claimed that the
use of light-ion beams  for investigating subthreshold $\bar{\mrm{p}}$
production should be useful  to verify theoretical  models.  Secondary
effects like  re-scattering, re-absorption,  self-energies, potentials
etc., which hide and   influence  features of the  primary  production
process, should be smaller than in heavy-ion reactions.  Indeed, it is
the  outstanding feature of the  data that  in d- and $\alpha$-induced
reactions  an enormous   enhancement of  $\bar{\mrm{p}}$~production by
nearly  two  and three  orders   of magnitude   compared to  p-induced
interactions  could be   observed.    The very fact  of   finding this
enhancement in light-ion induced reactions makes this interaction type
a   candidate for the key  to   a deeper understanding of subthreshold
$\bar{\mrm{p}}$~production in nucleus-nucleus reactions.

The KEK group~\cite{sugaya98} interprets their $\bar{\mrm{p}}$~data by
using  the `first-chance NN  collision model' from \cite{shor90} where
the internal nucleon  momenta  were  extracted from backward    proton
production  \cite{geaga80}    as  a  superposition of     two Gaussian
distributions.     In this  way,   the     momentum spectra and    the
incidence-energy   dependence   of  p-induced  reactions    could   be
successfully   reproduced by   adapting  one normalization  parameter.
However, the  application of this model  to  d-induced reactions using
different deuteron wavefunctions  and the normalization from p-induced
interactions    severely   underestimates  the   $\bar{\mrm{p}}$ cross
sections  at   subthreshold      energies.  Thus,     according     to
\cite{sugaya98}, the effect cannot be explained by the internal motion
of the nucleons in the deuteron,  even if a deuteron wavefunction with
a high-momentum component is used.

In   \cite{cassing94a},  $\bar{\mrm{p}}$~production   in  p+A  and d+A
reactions is analysed   within a phase-space  model incorporating  the
self-energies of the baryons.    It   is claimed that   the   internal
momentum distribution  of the deuteron  provides a natural explanation
of the large enhancement under discussion.

To the best of  our knowledge there is   no paper which  describes the
further  increase  of   subthreshold   $\bar{\mrm{p}}$~production   in
$\alpha$-induced reactions. Thus, the approach presented here seems to
be the first attempt  to consider  the whole set  of projectiles  in a
unified picture.  Beyond it, we also regard the whole set of ejectiles
measured.  This  is of special importance since  there is not only the
enhancement  of $\bar{\mrm{p}}$~production,  but  also the increase of
the $\mrm{K}^-$~cross sections  with increasing energy and mass number
of  the  projectile around  the    elementary production threshold  at
2.6~GeV. And also the completely different  energy and projectile mass
dependence  of the  pion  production  cross  sections  far   above the
threshold should be explained by a realistic approach.

In  figure~\ref{fCAll}, the results  of the ROC  model calculated with
one fixed parameter set are compared to the data~\cite{sugaya98}.  The
overall  agreement  is quite  satisfactory  in  view  of the different
projectile types, the large region of  incidence energies, the variety
of  ejectile  species  and the  huge   differences of many  orders  of
magnitude in the considered cross section values.  Particle yields are
influenced   by   the    suppression   factor         $\lambda_s=0.15$
(equation~(\ref{lamba})) of strange  quarks  and by the algorithm  for
creating the final hadrons from the quarks produced in the first stage
of    the  interaction process.   Hadrons   are   built-up in each  FB
independently        according     to    the      rules     of   quark
statistics~\cite{anisovich73}   by  randomly selecting  sequences   of
quarks  q and antiquarks  $\bar{\mrm{q}}$.  A q$\bar{\mrm{q}}$ gives a
meson,  while baryons  or    antibaryons  are  formed from  qqq     or
$\bar{\mrm{q}}\bar{\mrm{q}}\bar{\mrm{q}}$.  From a given sequence   of
quarks the different hadrons are formed according to the tables of the
particle data   group~\cite{PDG98}.    There  is no   parameter  which
directly determines the ratio  between meson and baryon production as,
e.g.,             in                the                    PYTHIA-LUND
model~\cite{andersson83,andersson87,sjostrand87a,pythia94}.    Only an
indirect influence via the  temperature  and the volume  parameters is
possible, which changes the relative  weights of the FBs in dependence
on their invariant mass and final particle multiplicity, respectively.
(See \cite{muellerh01} for  a  more detailed consideration   of hadron
formation.)

It  is,  therefore, remarkable that  the  general trend  of the energy
dependence  of the data  for all  projectile types, moderate  increase
with  energy of  the  cross sections for  the  light  mesons and steep
increase of  that  for  the antiprotons,   is  well described by   the
calculations.       Also  the    shift    of   the     maximum  in the
$\bar{\mrm{p}}$~spectra  towards   higher  momenta  is reproduced.   A
similar tendency can  be observed if  the cross  sections from dC  and
$\alpha$C reactions are  compared to those from  pC interactions.  The
heavier the observed  particle the steeper the  increase of the  cross
sections   for the heavier   projectiles,   especially at the   lowest
incidence energies.
\begin{figure}
\begin{center}
   \resizebox{0.7\textwidth}{!}{
   \includegraphics*{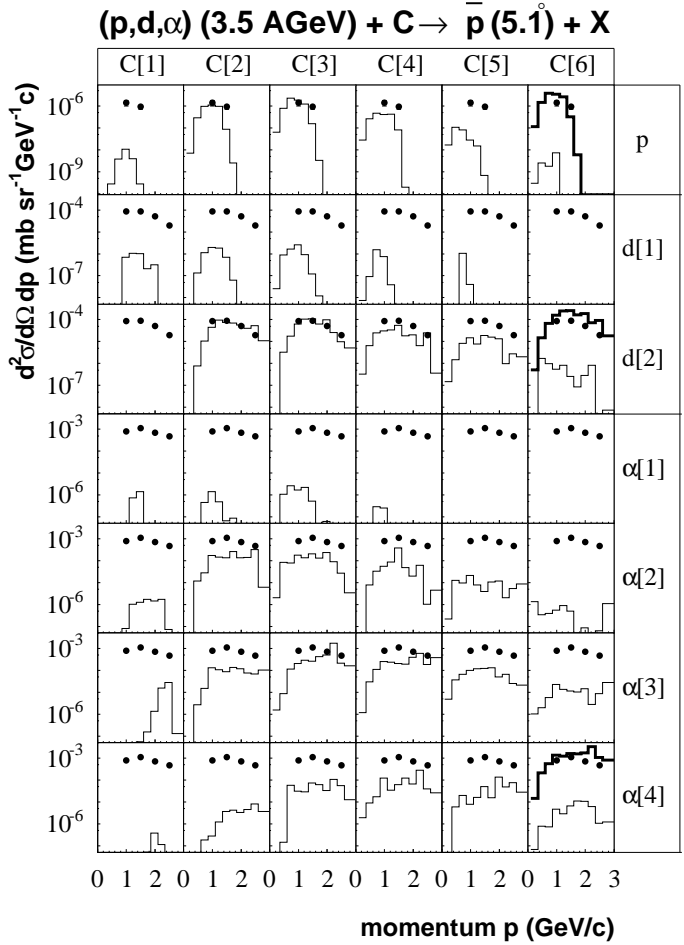}}
\end{center}
   \caption{\label{fCSub}           Partial            spectra      of
   $\bar{\mrm{p}}$~production calculated  for   pC, dC  and  $\alpha$C
   reactions (thin histograms)  compared  to the  data~\cite{sugaya98}
   (full circles).  In six columns denoted  by $C[b]$ the spectra from
   interactions  with  $b$ participating target  nucleons are plotted.
   Likewise, in the rows  the numbers of participants  in case of $d$-
   and  $\alpha$-induced   reactions  is    denoted  by  $d[a]$    and
   $\alpha[a]$,  respectively.  The thick histograms  are  the sums of
   the partial spectra and   coincide with the corresponding   results
   from figure~\ref{fCAll}.}
\end{figure}
\begin{figure}
\begin{center}
   \resizebox{0.7\textwidth}{!}{
   \includegraphics*{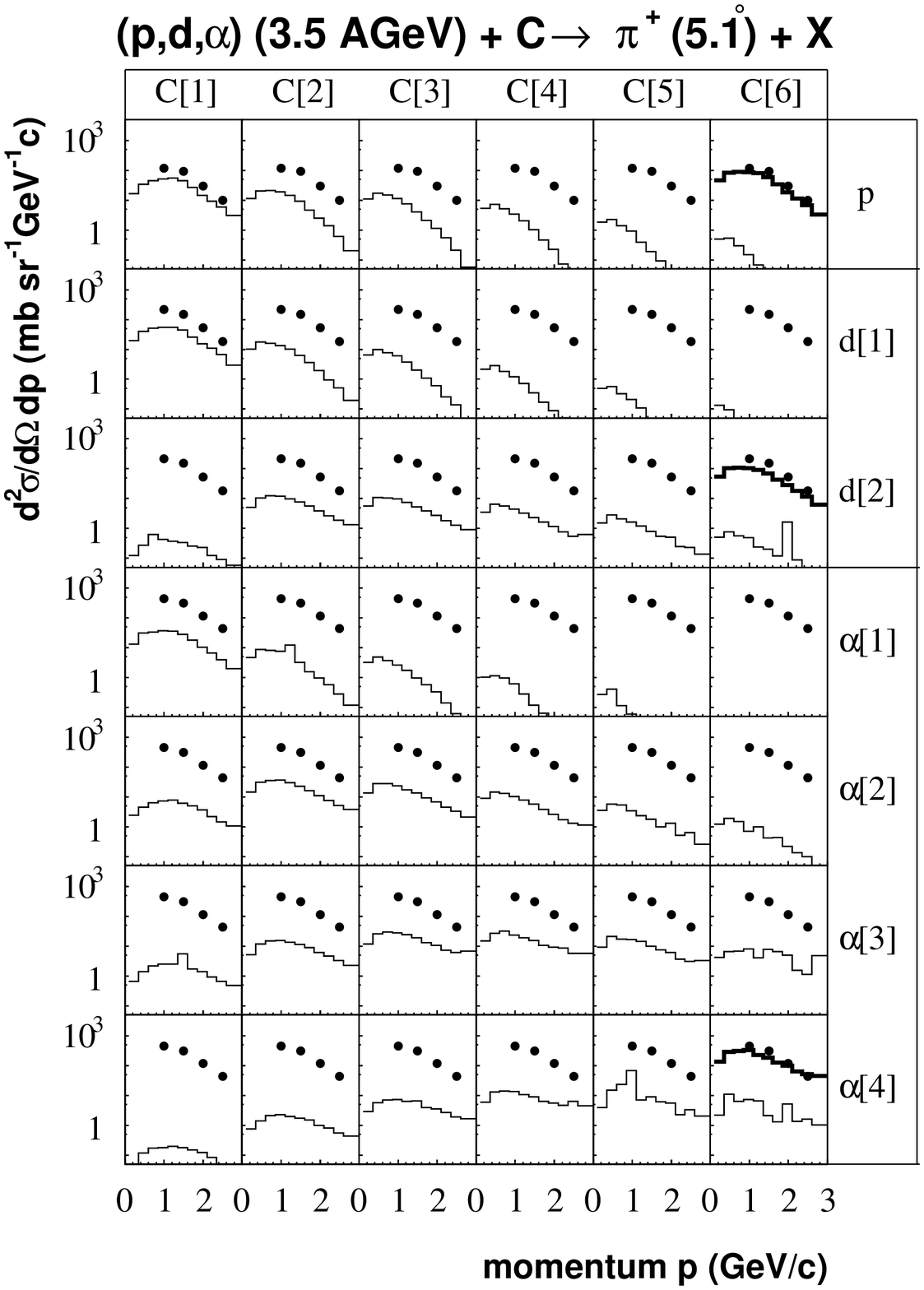}}
\end{center}
   \caption{\label{fCSubPi} The same as figure~\ref{fCSub} for $\pi^+$.}
\end{figure}

In  order to  make  the presented   results more  obvious  we  show in
figures~\ref{fCSub}   and \ref{fCSubPi}   the   partial spectra   from
collisions of $a$ projectile with $b$ target nucleons
\[
[a] + [b]
\]
for   the three   projectile    types   and two   selected   ejectiles
$\bar{\mrm{p}}$ and    $\pi^+$.  These   partial   spectra   represent
incoherent  sums over contributions  from  all possible charge numbers
(see (\ref{dsigma})) according to
\begin{eqnarray}
  \di\sigma_{ab}(s)               &               =                  &
  \sum_{z_a=max(0,a-\mc{N}_A)}^{min(a,\mc{Z}_A)}
  \sum_{z_b=max(0,b-\mc{N}_B)}^{min(b,\mc{Z}_B)}      \nonumber     \\
  &&\sigma_\ab        \,\frac{\di   W(s;\az)}{\sum_{\az}\int       \di
  W(s;\az)}\,\nonumber \,.
\end{eqnarray}
The partial  spectra result from  the interplay  between the values of
the integrated partial cross sections
\[
\sigma_{ab} = \sum_{z_a=max(0,a-\mc{N}_A)}^{min(a,\mc{Z}_A)}
\sum_{z_b=max(0,b-\mc{N}_B)}^{min(b,\mc{Z}_B)} \sigma_\ab\,.
\]
and the  energy  dependence  of  the relative weights   $\di W(s;\az)$
representing    the   production cross   section    in  the  $[a]+[b]$
interaction.  In the case of  p+A reactions the $\sigma_{1b}$ decrease
monotonically   with  increasing number  $b$   of participating target
nucleons, while for A+A  interactions symmetric combinations $a=b$ are
preferred.  Decisive for subthreshold particle production is the steep
increase  of  the production  cross section  with the available energy
above the threshold  in the $[a]+[b]$ interaction under consideration.
This explains the small contributions from the $[1]+[1]$ interactions,
although $\sigma_{11}$ represents for all projectile types the largest
partial cross  section.    It  is the small  available    energy which
prevents higher $\bar{\mrm{p}}$ production.  With increasing number of
participants both   the  mean  available  energy   and the  number  of
competing   final  channels    increase.  This   results    in maximal
contributions from the combinations $[1]+[2]$ and $[1]+[3]$ in case of
p+A reactions.  The spectral  distributions are rather independent  of
the number   of     target  participants  $b$,  see    first  row   in
figure~\ref{fCSub}.   This explains the  successful description of the
spectra in the `first-chance NN collision' approach, which corresponds
to the $[1]+[1]$ process of the ROC  model, by appropriately selecting
the normalization \cite{sugaya98}.  Thus, the interpretation of the pC
data by the  ROC model is quite  contradictory to the assumptions of a
`first-chance NN collision model'.    Comparing the $[1]+[1]$  spectra
from  pC and dC reactions only  a moderate increase of $\bar{\mrm{p}}$
production is predicted by  the ROC model due  to the Fermi motion  in
the  deuteron.  The enormous enhancement   observed stems mainly  from
processes  with both  nucleons of  the   deuteron  and several  target
nucleons  involved where the  energy available for particle production
is  much larger than  in  $[1]+[1]$ processes.   In  case of $\alpha$C
reactions,  the  number of   partial  spectra contributing essentially
increases.   Here the larger available    energy arises both from  the
larger masses  of the interacting  subsystems and from  their internal
motion in the projectile and the target.   Thus, one can conclude from
the viewpoint of the ROC model that the key quantity for understanding
subthreshold   particle  production is  the   number  of participating
nucleons.  A direct   experimental determination  of  this number  for
subthreshold   $\bar{\mrm{p}}$~production   is   highly   desirable as
discussed in \cite{muellerh96}  for the case of $\mrm{K}^-$~production
in pA reactions.

If  we consider  particle production   far above the  threshold  as in
figure~\ref{fCSubPi}  then   the  main   difference  to   subthreshold
production consists  in the higher  value of the energy  available for
the production  of the considered  particle.  This leads to a smoother
energy dependence of the  production  cross section for  the  particle
under    consideration and  the   gain  of   available  energy due  to
cluster-cluster processes does not play   such an important role.   In
this way,   the partial spectra  reflect  primarily the  values of the
corresponding partial  cross  sections $\sigma_{ab}$   with only minor
corrections  due to  the energy  dependence   of the production  cross
section.  Thus, the $[1]+[1]$ processes with the largest partial cross
section for all projectile types yield also  the main contributions to
the spectra.

\section{Summary \label{sum}}

Subthreshold particle production is  a collective phenomenon  which is
far from  being completely understood.  From  the viewpoint of the ROC
model, data on subthreshold particle  production can be reproduced  by
considering the interaction  of few-nucleon groups in complete analogy
to   the    interaction  of  single nucleons,   also    with regard to
high-momentum transfer processes.   It has been demonstrated here that
the   cluster concept yields   a  quite   natural explanation  of  the
enhancement of subthreshold particle production due to the energy gain
in the interaction of  few-nucleon groups compared to NN interactions.
This   concept   should be    applicable  not   only   in proton-   or
light-ion-induced   reactions, but also   for heavy-ion  interactions,
although in the latter case  the number of partial processes increases
tremendously.   In this  sense, the ROC  model  can be considered as a
promising approach to   a unified description of particle   production
processes in a large variety of different types of nuclear reactions.

\ack
One of the  authors (H.M.) would like  to thank  W.~Eng\-hardt for the
promotion of this study and A.~Sibirtsev for useful discussions.

\section*{References}
\bibliographystyle{JourPhysG}
\bibliography{COSY,hadMod,hmbib,hmhelp,nucData,frag,Monte_Carlo}

\begin{thebibliography}{10}
\expandafter\ifx\csname url\endcsname\relax
  \def\url#1{\texttt{#1}}\fi
\expandafter\ifx\csname urlprefix\endcsname\relax\def\urlprefix{URL }\fi
\providecommand{\eprint}[2][]{\url{#2}}

\bibitem{chamberlain55}
Chamberlain O, Segr{\'{e}} E, Wiegand C and Ypsilantis T 1955 \emph{Phys. Rev.}
  \textbf{100} 947

\bibitem{chamberlain56}
Chamberlain O, Chupp W, Goldhaber G \emph{et~al.} 1956 \emph{Nuovo Cimento}
  \textbf{3} 447

\bibitem{elioff62}
Elioff T \emph{et~al.} 1962 \emph{Phys. Rev.} \textbf{128} 869

\bibitem{dorfan65a}
Dorfan D~E \emph{et~al.} 1965 \emph{Phys. Rev. Lett.} \textbf{14} 995

\bibitem{baldin88}
Baldin A \emph{et~al.} 1988 \emph{JETP Lett.} \textbf{48} 137

\bibitem{caroll89}
Caroll J \emph{et~al.} 1989 \emph{Phys. Rev. Lett.} \textbf{62} 1829

\bibitem{shor89}
Shor A \emph{et~al.} 1989 \emph{Phys. Rev. Lett.} \textbf{63} 2192

\bibitem{schroeter93}
Schr\"oter A \emph{et~al.} 1993 \emph{Nucl. Phys. A} \textbf{553} 775c

\bibitem{schroeter94}
Schr\"oter A \emph{et~al.} 1994 \emph{Z. Phys. A} \textbf{350} 101

\bibitem{lepikhin87}
Lepikhin Y~B, Smirnitsky V~A and Sheinkman V~A 1987 \emph{JETP Lett.}
  \textbf{46} 275

\bibitem{chiba93}
Chiba J \emph{et~al.} 1993 \emph{Nucl. Phys. A} \textbf{553} 771c

\bibitem{sugaya98}
Sugaya Y \emph{et~al.} 1998 \emph{Nucl. Phys.} \textbf{A634} 115

\bibitem{li94a}
Li G~Q, Ko C~M, Fang X~S and Zheng Y~M 1994 \emph{Phys. Rev. C} \textbf{49}
  1139

\bibitem{batko91}
Batko G, Cassing W, Mosel U, Niita K and Wolf G 1991 \emph{Phys. Lett. B}
  \textbf{256} 331

\bibitem{huang92}
Huang S~W, Li G, Maruyama T and Faessler A 1992 \emph{Nucl. Phys. A}
  \textbf{547} 653

\bibitem{cassing92}
Cassing W, Lang A, Teis S and Weber K 1994 \emph{Nucl. Phys. A} \textbf{545}
  123c

\bibitem{teis93}
Teis S, Cassing W, Maruyama T and Mosel U 1993 \emph{Phys. Lett. B}
  \textbf{319} 47

\bibitem{teis94}
Teis S, Cassing W, Maruyama T and Mosel U 1994 \emph{Phys. Rev. C} \textbf{50}
  388

\bibitem{cassing94a}
Cassing W, Lykasov G and Teis S 1994 \emph{Z. Phys. A} \textbf{348} 247

\bibitem{batko94}
Batko G, Faessler A, Huang S, Lehmann E and Rajeev K 1994 \emph{J. Phys. G}
  \textbf{20} 461

\bibitem{hernandez95}
Hern{\'a}ndez E, Oset E and Weise W 1995 \emph{Z. Phys. A} \textbf{351} 99

\bibitem{sibirtsev98}
Sibirtsev A, Cassing W, Lykasov G~I and Rzjanin M~V 1998 \emph{Nucl. Phys. A}
  \textbf{632} 131

\bibitem{koch89}
Koch P and Dover C~B 1989 \emph{Phys. Rev. C} \textbf{40} 145

\bibitem{ko88}
Ko C~M and Ge X 1988 \emph{Phys. Lett. B} \textbf{205} 195

\bibitem{ko89}
Ko C~M and Xia L~H 1989 \emph{Phys. Rev. C} \textbf{40} R1118

\bibitem{dyachenko00}
D'yachenko A~T 2000 \emph{J. Phys. G} \textbf{26} 861

\bibitem{danielewicz90}
Danielewicz P 1990 \emph{Phys. Rev. C} \textbf{42} 1564

\bibitem{Komarov04}
Komarov V~I, M\"uller H and Sibirtsev A 2004 \emph{J. Phys. G.} \textbf{30}
  921.
\newblock \eprint[http://arXiv.org]{http://arXiv.org/nucl-th/0312087}

\bibitem{muellerh95}
M\"uller H 1995 \emph{Z. Phys. A} \textbf{353} 103

\bibitem{muellerh01}
M\"uller H 2001 \emph{Eur. Phys. J. C} \textbf{18} 563.
\newblock \eprint[http://arXiv.org]{http://arXiv.org/hep-ph/0011350}

\bibitem{muellerh92}
M\"uller H and Sistemich K 1992 \emph{Z. Phys. A} \textbf{344} 197

\bibitem{muellerh91}
M\"uller H 1991 \emph{Z. Phys. A} \textbf{339} 409

\bibitem{muellerh95a}
M\"uller H 1995 \emph{Z. Phys. A} \textbf{353} 237

\bibitem{muellerh96}
M\"uller H 1996 \emph{Z. Phys. A} \textbf{355} 223

\bibitem{shmakov88}
Shmakov S, Uzhinskii V and Zadorozhny A 1988 \emph{Comp. Phys. Communications}
  \textbf{54} 125

\bibitem{glauber70}
Glauber R~J and Mathiae J 1970 \emph{Nucl. Phys. B} \textbf{21} 135

\bibitem{knoll79}
Knoll J 1979 \emph{Phys. Rev. C} \textbf{20} 773

\bibitem{knoll80}
Knoll J 1980 \emph{Nucl. Phys. A} \textbf{343} 511

\bibitem{bohrmann81}
Bohrmann S and Knoll J 1981 \emph{Nucl. Phys. A} \textbf{356} 498

\bibitem{shyam84}
Shyam R and Knoll J 1984 \emph{Nucl. Phys. A} \textbf{426} 606

\bibitem{shyam86}
Shyam R and Knoll J 1986 \emph{Nucl. Phys. A} \textbf{448} 322

\bibitem{knoll88}
Knoll J and Shyam R 1988 \emph{Nucl. Phys. A} \textbf{483} 711

\bibitem{ghosh90}
Ghosh B and Shyam R 1990 \emph{Phys. Lett. B} \textbf{234} 248

\bibitem{ghosh92}
Ghosh B 1992 \emph{Phys. Rev. C} \textbf{45} R518

\bibitem{elton61}
Elton L~R 1961 \emph{Nuclear sizes/L. R. B. Elton} (London : Oxford Univ. Pr.)

\bibitem{lacombe81}
Lacombe M \emph{et~al.} 1981 \emph{Phys. Lett. B} \textbf{101} 139

\bibitem{goldhaber74}
Goldhaber A~S 1974 \emph{Phys. Lett. B} \textbf{53} 306

\bibitem{shor90}
Shor A, Perez-Mendez V and Ganezer K 1990 \emph{Nucl. Phys. A} \textbf{514} 717

\bibitem{geaga80}
Geaga J~V \emph{et~al.} 1980 \emph{Phys. Rev. Lett.} \textbf{45} 1993

\bibitem{ciofi96}
{Ciofi degli Atti} C and Simula S 1996 \emph{Phys. Rev. C} \textbf{53} 1689

\bibitem{benhar89}
Benhar O, Fabrocini A and Fantoni S 1989 \emph{Nucl. Phys. A} \textbf{505} 267

\bibitem{sick94}
Sick I, Fantoni S, Fabrocini A and Benhar O 1994 \emph{Phys. Lett. B}
  \textbf{323} 267

\bibitem{sibirtsev97}
Sibirtsev A, Cassing W and Mosel U 1997 \emph{Z. Phys. A} \textbf{358} 357

\bibitem{byckling73}
Byckling E and Kajantie K 1973 \emph{Particle Kinematics} (John Wiley and Sons,
  London, New York, Sydney, Toronto)

\bibitem{westfall78}
Westfall G \emph{et~al.} 1978 \emph{Phys. Rev.} \textbf{C17} 1368

\bibitem{anisovich73}
Anisovich V~V and Shekhter V~M 1973 \emph{Nucl. Phys. B} \textbf{55} 455

\bibitem{PDG98}
Caso C \emph{et~al.} 1998 \emph{Eur. Phys. J. C} \textbf{3} 1

\bibitem{andersson83}
Andersson B, Gustafson G, Ingelman G and Sjostrand T 1983 \emph{Phys. Rep.}
  \textbf{97} 31

\bibitem{andersson87}
Andersson B, Gustafson G and Nilsson-Almqvist B 1987 \emph{Nucl. Phys. B}
  \textbf{281} 289

\bibitem{sjostrand87a}
Sj{\"o}strand T and van Zijl M 1987 \emph{Phys. Rev. D} \textbf{36} 2019

\bibitem{pythia94}
Sj\"ostrand T 1994 \emph{Comput. Phys. Commun.} \textbf{82} 74

\end{thebibliography}

\end{document}